\documentclass[journal,twoside]{IEEEtran}
\IEEEoverridecommandlockouts
\usepackage{cite}
\usepackage{amsmath,amssymb,amsfonts}
\usepackage{algorithmic}
\usepackage{graphicx}
\usepackage{textcomp}
\usepackage{xcolor}
\usepackage{orcidlink}

\def\BibTeX{{\rm B\kern-.05em{\sc i\kern-.025em b}\kern-.08em
    T\kern-.1667em\lower.7ex\hbox{E}\kern-.125emX}}
\begin{document}

\title{Reed-Muller Error-Correction Code Encoder for SFQ-to-CMOS Interface Circuits\\
}
\markboth{IEEE Transactions on Applied Superconductivity, Vol. 36, No. Y, Month 2026}
{Mustafa \MakeLowercase{\textit{et al.}}: Reed-Muller Error-Correction Code Encoder for SFQ-to-CMOS Interface Circuits}

\author{Yerzhan~Mustafa\orcidlink{0000-0001-7755-1626},~\IEEEmembership{Member,~IEEE,}
Berker Peköz\orcidlink{0000-0002-7572-3663},~\IEEEmembership{Member,~IEEE,}
        and~Selçuk~Köse\orcidlink{0000-0001-8095-6691},~\IEEEmembership{Member,~IEEE}
\thanks{Y. Mustafa and S. Köse are with the Department
of Electrical and Computer Engineering, University of Rochester, Rochester,
NY, 14627, USA. E-mails: (yerzhan.mustafa@rochester.edu, selcuk.kose@rochester.edu).

B. Peköz is with the Department of Electrical Engineering and Computer Science, Embry-Riddle Aeronautical University, Daytona Beach, FL, 32114, USA. E-mail: (berker.pekoz@erau.edu)}

\thanks{
The code and data used to support the findings of this study are available from the corresponding author upon reasonable request.

This work is supported in part by the National Science Foundation Expeditions Award under Grant CCF-2124453 and SHF Award under Grant CCF-2308863, and Department of Energy EXPRESS program under Grant DE-SC0024198.}
}

\maketitle

\begin{abstract}
Data transmission from superconducting digital electronics such as single flux quantum (SFQ) logic to semiconductor (CMOS) circuits is subject to bit errors due to, e.g., flux trapping, process parameter variations (PPV), and fabrication defects. In this paper, a lightweight hardware-efficient error-correction code encoder is designed and analyzed. Particularly, a Reed-Muller code RM(1,3) encoder is implemented with SFQ digital logic. 
The proposed RM(1,3) encoder converts a 4-bit message into an 8-bit codeword and can detect and correct up to 3- and 1-bit errors, respectively. 
This encoder circuit is designed using MIT-LL SFQ5ee process and SuperTools/ColdFlux RSFQ cell library. 
A simulation framework integrating JoSIM simulator and MATLAB script for automated data collection and analysis, is proposed to study the performance of RM(1,3) encoder. The proposed encoder improves the probability of having no bit errors by 6.7\% as compared to an encoder-less design under $\pm$20\% PPV.
With $\pm$15\% and lower PPV, the proposed encoder could correct all errors with at least 99.1\% probability. 
The impact of fabrication defects such as open circuit faults on the encoder circuit is also studied using the proposed framework.
\end{abstract}


\section{Introduction}
\IEEEPARstart{S}{ingle} flux quantum logic (SFQ) is a type of superconducting digital electronics that can operate at extremely high switching frequencies (tens to hundreds of GHz) and consume significantly low energy per switching activity, in the order of 10$^{-19}$~J \cite{likharev1991rsfq,krylov2024single}. 
SFQ technology is a promising candidate for beyond-CMOS applications, such as data centers and cloud computing. Additionally, it can be used in large-scale superconducting quantum computers as in-fridge control and readout circuitry \cite{mukhanov2019scalable,jokar2022digiq}. 

In SFQ logic, the information is represented in terms of voltage pulses, which are typically around 1 mV in amplitude with 2 ps duration \cite{likharev1991rsfq}.
Since semiconductor electronics (\textit{e.g.}, CMOS) operate with DC signal levels from a few hundred millivolts (mV) up to over 1 V, special SFQ-to-CMOS interface circuits are required to convert and amplify SFQ pulses.
Various types of superconducting output drivers are available in the literature, such as Suzuki stack \cite{suzuki1988josephson,ortlepp2013design,mustafa2023suzuki,hironaka2025josephson,mustafa2024ternary}, SQUID (superconducting quantum interference device) stack \cite{gupta2019digital,razmkhah2021compact,egan2022true,zhao2025high,mustafa2025pam4_squid}, and SFQ-to-DC converter \cite{kaplunenko1989experimental,likharev1991rsfq,ortlepp2009superconductor,inamdar2009superconducting}. These interface circuits and standard SFQ logic cells are susceptible to flux trapping \cite{robertazzi1997flux,fourie2021experimental}, process parameter variations (PPV) \cite{tolpygo2014inductance}, and fabrication defects such as open/short circuits \cite{mustafa2024built}. Each of these non-ideal effects may cause bit errors during the data transmission from the SFQ chip operating at 4.2~K to a higher temperature stage (\textit{e.g.}, 50-300 K). These mechanisms can create both isolated bit errors and occasional errors bursts  along the transmitted data stream. Therefore, instead of assuming a fixed, time-invariant bit-error rate, we evaluate the error statistics directly through circuit-level simulations under PPV and defect variations.

\begin{figure*}[t]
	\centering
	\includegraphics[width=0.98\textwidth]{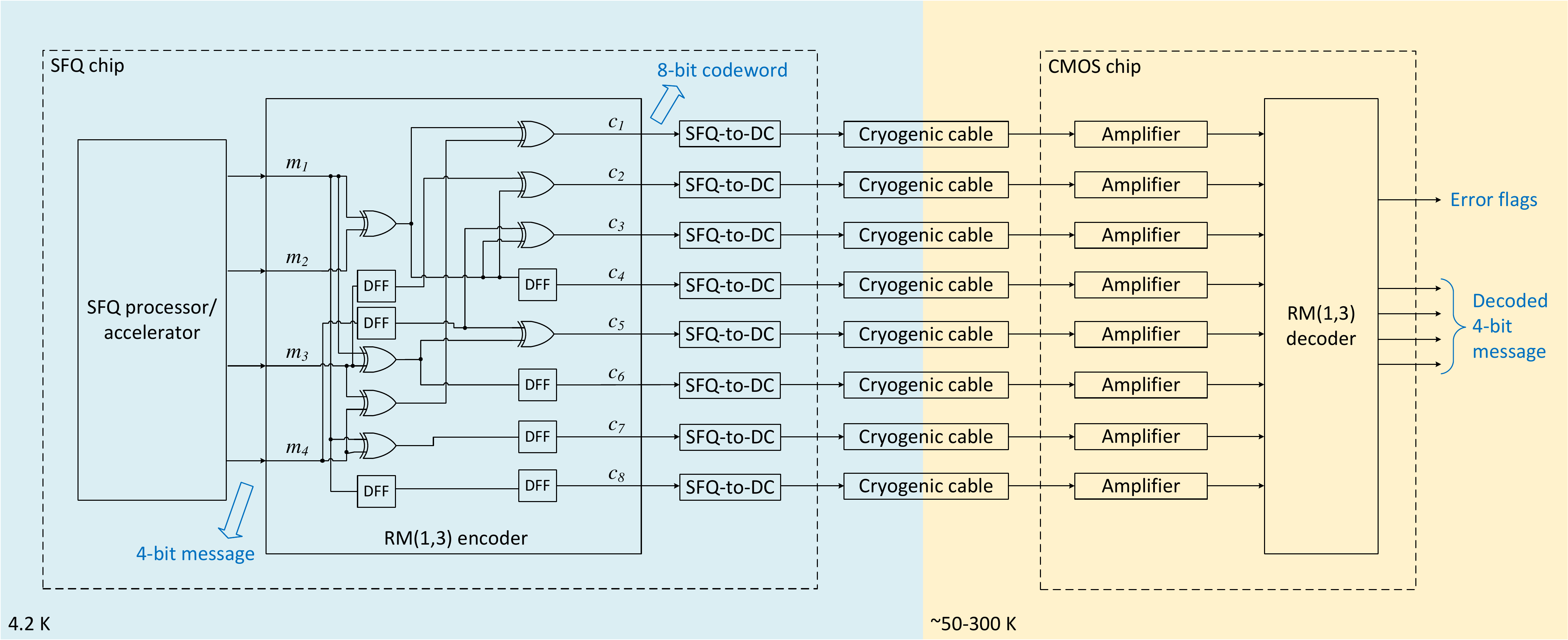}
	\caption{Block diagram of a cryogenic digital output data link incorporating the SFQ-based RM(1,3) encoder and decoder. The schematic of RM(1,3) encoder is shown at the logic level using XOR gates, DFFs, and SFQ splitters (clock distribution network is not shown).}
	\label{fig:RM13_schematic_in_data_link}
\end{figure*}

Error-correction codes (ECCs) are used to detect and correct bit errors during data transmission and storage\cite{clemente2022reliability}. Although a more complex ECC may detect and correct a larger number of bit errors, it often requires more hardware resources (\textit{i.e.}, parity bits and encoder logic). Within cryogenic environment, the cooling power and area constraints are key components that should be considered during the selection of ECCs. 
Having additional parity bits require a larger number of cryogenic cables, whose heat load is a major component {of} the cooling power budget \cite{krinner2019engineering,mustafa2024dc}. 

An SFQ-based (38,32) linear block code encoder has been presented in \cite{peng2019solution}. 
This encoder can transmit a 32-bit message and six parity bits. While this encoder can detect 2-bit and correct 1-bit errors, it requires 84 XOR gates and 135 D flip-flops (DFFs) \cite{peng2019solution}. 
%
Given these constraints, we focus on a short-block linear code that (i) can correct single-bit errors and detect multiple-bit errors within each codeword, and (ii) admits a low-latency implementation compatible with high-data-rate of SFQ logic. In this work, we therefore adopt the first-order RM(1,3) code, which is an [8,4,4] linear block code with minimum Hamming distance 4. This structure enables correction of one bit error and detection of up to three bit errors in each 8-bit codeword encoding 4-bit messages, while maintaining a simple XOR-based encoder architecture.

The following are the key contributions of this work. 

\begin{itemize}
    \item A lightweight hardware-efficient ECC encoder is designed with SFQ logic using Reed-Muller (RM) code. 

    \item A simulation framework, which co-integrates JoSIM simulator \cite{delport2019josim} and MATLAB tool, is proposed for automated data collection and analysis. 

    \item Using the proposed framework, the effects of PPV and fabrication defects on the performance of encoder are studied. 
\end{itemize}

The rest of the paper is organized as follows. In Section \ref{section:RM_codes}, the RM codes are briefly overviewed and an encoder circuit is designed with SFQ logic cells. The proposed automated simulation framework is proposed in Section \ref{section:framework} with the performance analysis of SFQ-based lightweight encoder. Conclusions are drawn in Section \ref{conclusion}.

\section{Reed-Muller (RM) Codes}\label{section:RM_codes}
The RM codes are overviewed in Section \ref{section:RM_overview}. The circuit-level design of RM(1,3) code encoder, which is implemented with SFQ logic, is presented in Section \ref{section:RM_encoder_circuit}.

\subsection{Overview}\label{section:RM_overview}

RM codes were independently introduced by Irving Reed\cite{reed1954class} and David Muller\cite{muller1954application} in 1954. RM codes are linear error-correction block codes that are commonly used when high reliability is needed, such as tele- and satellite communication, data storage and military/aerospace secure links \cite{abberm}. Two parameters, $r$ and $m$, define the message length ($k = \sum_{i=0}^{r} {m\choose i}$) and the block length ($2^m$) of RM($r$,$m$) code.
%
%



\subsection{Circuit-level design of RM(1,3) encoder with SFQ logic}\label{section:RM_encoder_circuit}

An RM code can be represented with a generator matrix $G$, which can convert the message to the codeword as follows

\begin{equation}
    codeword = (message\times G)\mod 2.
\label{eq:codeword_equation}
\end{equation}
\noindent
For example, RM(1,3) code has a generator matrix 

\begin{equation}
    G_{RM(1,3)} = 
\begin{bmatrix}
    1 & 1 & 1 & 1 & 1 & 1 & 1 & 1\\
    1 & 1 & 1 & 1 & 0 & 0 & 0 & 0\\
    1 & 1 & 0 & 0 & 1 & 1 & 0 & 0\\
    1 & 0 & 1 & 0 & 1 & 0 & 1 & 0
\end{bmatrix}.
\label{eq:generator_matrix_RM13}
\end{equation}
\noindent
Using (\ref{eq:codeword_equation}), each bit of the codeword can be expressed as

\begin{equation}
\begin{array}{l}
    c_1 = m_1 \oplus m_2 \oplus m_3 \oplus m_4;\\
    c_2 = m_1 \oplus m_2 \oplus m_3;\\
    c_3 = m_1 \oplus m_2 \oplus m_4;\\
    c_4 = m_1 \oplus m_2;\\
    c_5 = m_1 \oplus m_3 \oplus m_4;\\
    c_6 = m_1 \oplus m_3;\\
    c_7 = m_1 \oplus m_4;\\
    c_8 = m_1,
\end{array}
\label{eq:codeword_boolean}
\end{equation}
\noindent
where $\oplus$ is an XOR operator, $message=[m_1,m_2,m_3,m_4]$, and $codeword = [c_1,c_2,...,c_8]$. 

To convert the Boolean representation (\ref{eq:codeword_boolean}) to a hardware-efficient circuit-level design, the unique features of SFQ logic must be considered, as listed below. 

\begin{enumerate}
    \item Unlike CMOS logic, the layout area and power consumption of a 2-input SFQ XOR gate are similar to those of 2-input AND or OR gates \cite{supertools_rsfq_cell_library}. 

    \item SFQ logic cells have a single fan-out. Hence, an SFQ splitter circuit is required whenever a gate drives more than one SFQ cell.

    \item Each SFQ logic cell such as NOT, AND, OR, and XOR requires a clock signal to propagate the output signals. Consequently, each data path must be balanced to ensure proper signal arrival at each gate. Path balancing is typically achieved by adding DFF cells~\cite{pasandi2018pbmap}.

\end{enumerate}
The schematic of SFQ-based RM(1,3) code encoder is depicted in Fig.~\ref{fig:RM13_schematic_in_data_link}. The encoder circuit consists of 8 XOR gates, 7 DFFs, and 26 splitters (14 of them are used to form a clock distribution network). 
Together with the 8 SFQ-to-DC converters (Fig.~\ref{fig:RM13_schematic_in_data_link}), the encoder circuit dissipates 101.5 $\mu$W and occupies 0.193 mm$^2$ of layout area. These estimates are based on SuperTools/ColdFlux RSFQ cell library \cite{supertools_rsfq_cell_library} with MIT Lincoln Lab SFQ5ee 10~kA/cm$^2$ process. 

\begin{figure}[t]
	\centering
	\includegraphics[width=0.5\textwidth]{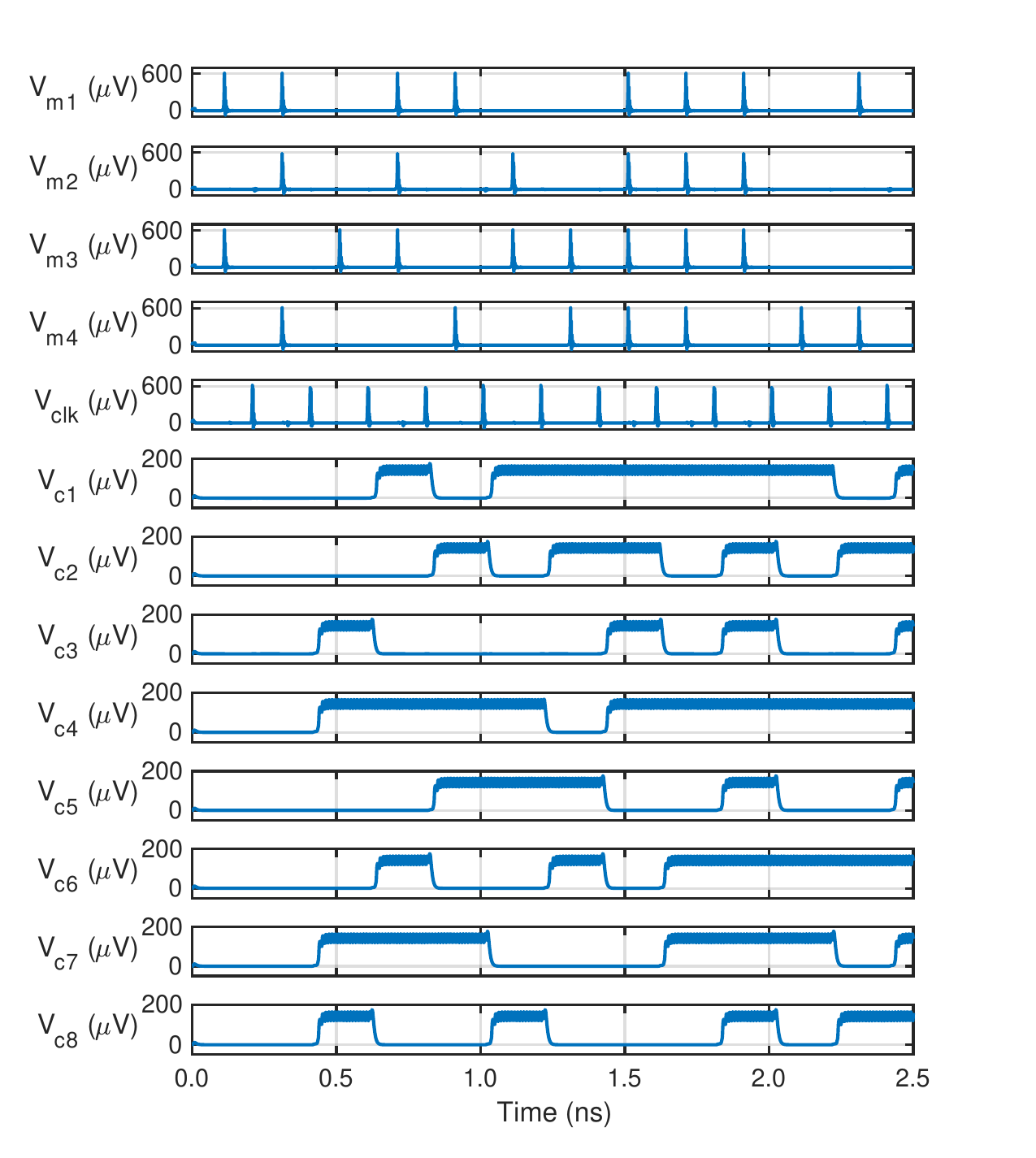}
	\caption{Simulation results of the SFQ-based RM(1,3) code encoder operating at 5~GHz. SFQ-to-DC converters are used as an interface circuit.}
	\label{fig:RM13_waveforms}
\end{figure}

The simulation results of SFQ-based RM(1,3) code encoder are shown in Fig. \ref{fig:RM13_waveforms}. The circuit is simulated in JoSIM \cite{delport2019josim} with SuperTools/ColdFlux RSFQ cell library \cite{supertools_rsfq_cell_library}. As can be observed from Fig. \ref{fig:RM13_waveforms}, it takes two clock cycles to produce the codeword bits. For instance, the first message bits `1010' are applied at 0.1 ns, the codeword bits `00110011' are produced at 0.4 ns as shown in Fig. \ref{fig:RM13_waveforms}. It should be noted that SFQ-to-DC converters are used as in interface circuits, which produce DC voltage signals with non-return-to-zero (NRZ) scheme. Particularly, a logical `1' is observed when the output voltage of SFQ-to-DC converter transitions from high-to-low or low-to-high \cite{mustafa2024built}. Similarly, a logical `0' corresponds to a no transition in the output voltage.

\section{Proposed Simulation Framework for Performance Analysis of SFQ-based Encoders}\label{section:framework}

JoSIM offers useful built-in functions such as circuit parameter spread (variation) and thermal noise insertion \cite{delport2019josim}. However, sophisticated fault analysis of SFQ circuits requires external tools to expand JoSIM's standard functionality. In this section, a MATLAB tool is used with JoSIM to develop a simulation framework for automated ECC data collection and analysis. A block diagram of the proposed framework is shown in Fig.~\ref{fig:framework_v2}.
In this figure, a custom MATLAB script is written that performs the following actions:

\begin{figure}[t]
	\centering
	\includegraphics[width=0.5\textwidth]{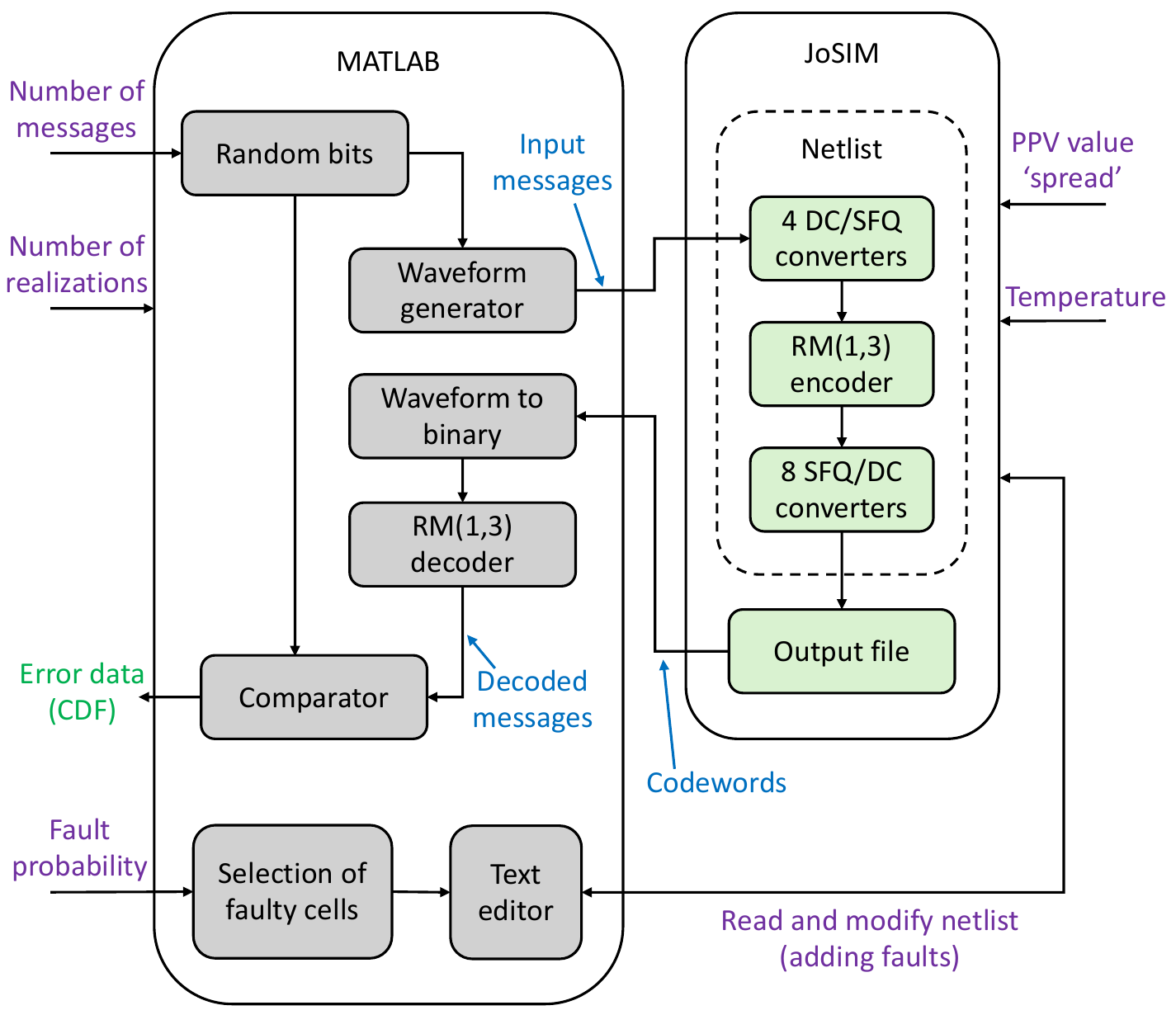}
	\caption{Block diagram of the proposed simulation framework for the analysis of RM(1,3) code encoder.}
	\label{fig:framework_v2}
\end{figure}

\begin{enumerate}
    \item Generate a set of random messages in a binary format. 

    \item Convert these binary messages into continuous waveforms, which will be fed to the input terminal of DC-to-SFQ converter circuits. These waveforms are saved as a text file and are used as the custom waveform `\texttt{cus()}' function in JoSIM netlist.

    \item Execute a standard JoSIM command line, which performs the simulation, using `\texttt{system()}' function in MATLAB. 

    \item Read the output file of JoSIM, which is saved as a spreadsheet and contains the waveforms of SFQ-to-DC converter circuits, and convert it into a binary format. 

    \item Decode the received binary codewords using `\texttt{rmdec\_reed()}' function in \cite{RM_matlab_code}. 

    \item Compare the decoded messages with the original messages and plot a cumulative distribution function (CDF) using `\texttt{cdfplot()}' function in MATLAB. 
    
\end{enumerate}

The RM(1,3) decoder implemented with `\texttt{rmdec\_reed()}' function uses a majority-logic algorithm. Although, it can correct 1-bit errors in a 8-bit codeword, it is still susceptible to false positive decoding. During the performance analysis, the reliability of RM(1,3) decoder is also included by comparing the original and decoded messages as shown in Fig. \ref{fig:framework_v2}.

To analyze the performance of SFQ-based encoder circuits, non-ideal effects should be inserted into simulation that could cause bit errors. Particularly, the effects of PPV and fabrication defects are analyzed in Sections \ref{section:PPV_effects} and \ref{section:open_faults}. 
During the performance analysis of above-mentioned sources of errors, the interface circuits (particularly, SFQ-to-DC converters) are set to operate at 5 GHz, producing a total of 40 Gbps data rate across all channels. Although SFQ-to-DC converters are capable of operating at a higher speed of 30-40 Gbps per channel \cite{gupta2019digital,supertools_rsfq_cell_library}, a lower data rate is selected to reduce the errors from ``Waveform to binary" conversion (see Fig. \ref{fig:framework_v2}) and provide more period of time for signal averaging (\textit{i.e.}, higher signal-to-noise ratio) and binary classification. This approach enables detailed study of PPV and fabrication defects, which may cause permanent circuit failure irrespective of operating frequency. As a future research, the proposed simulation framework can be expanded with high data rate interface incorporating signal integrity simulations \cite{krause2024signal}, which can model the attenuation and noise from cryogenic interconnect and semiconductor amplifiers.

\subsection{Effects of PPV}\label{section:PPV_effects}

PPV effects can be modeled in JoSIM by using the `\texttt{.spread}' function. This function assigns a random value (within a specified range from the nominal value) for circuit parameters such as inductance, resistance, and critical current of JJ from a uniform distribution. With a defined spread parameter in JoSIM netlist, each simulation run has a specific set of circuit parameters. 

\begin{figure}[t]
	\centering
	\includegraphics[width=0.5\textwidth]{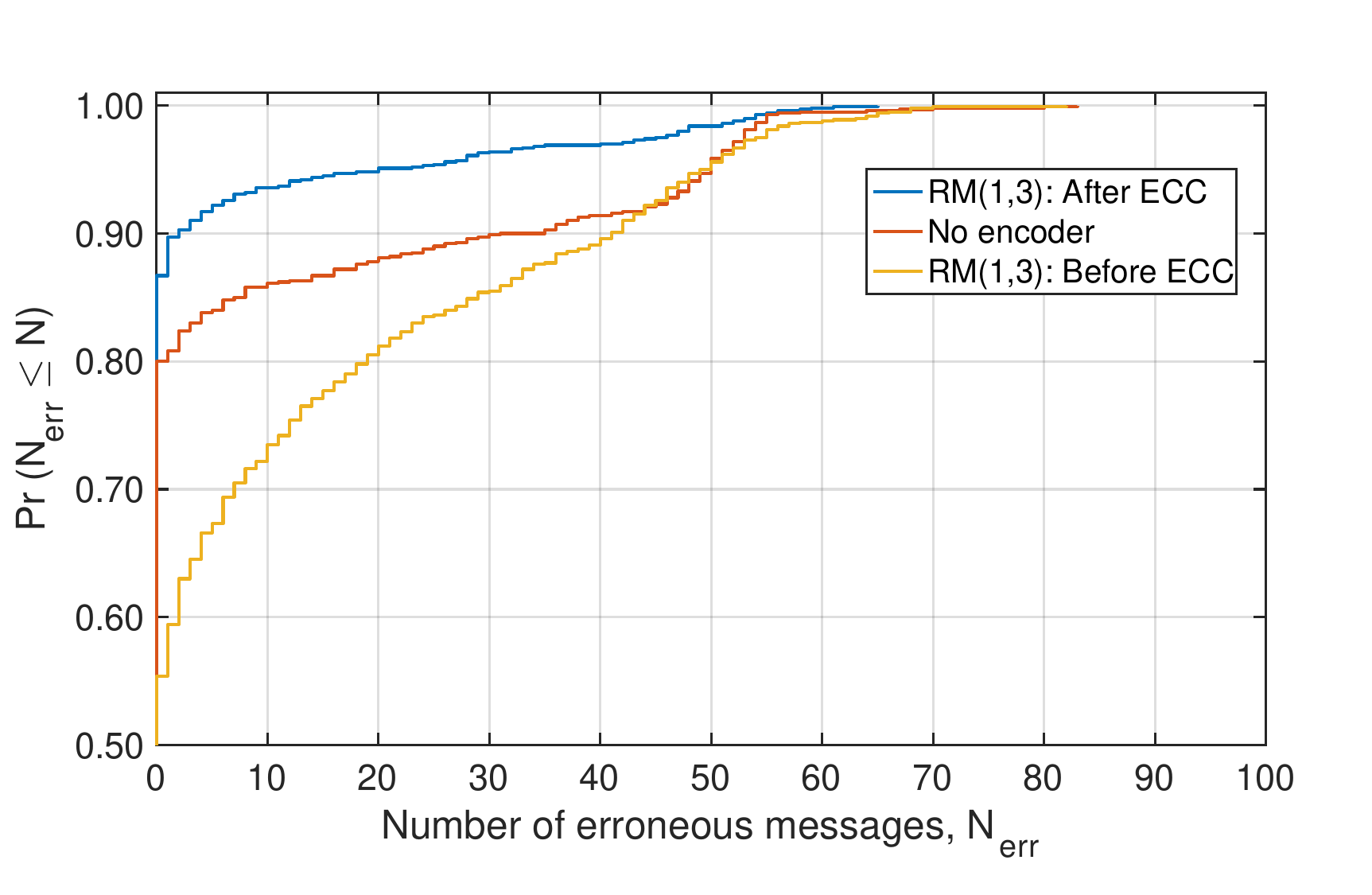}
	\caption{CDF for receiving at most {$N_\mathtt{err}$} erroneous messages. {$N=100$} messages are transmitted with 1000 independent circuit realizations. Each realization has up to $\pm$20\% PPV.}
	\label{fig:CDF_comparison_20PPV_v2}
\end{figure}

To analyze PPV effects, the `Number of realizations' parameter is introduced into the simulation framework (Fig.~\ref{fig:framework_v2}). This parameter sets the number of times JoSIM netlist is generated, where each realization can be viewed as a distinct fabricated chip. Fig. \ref{fig:CDF_comparison_20PPV_v2} shows the CDF plot for receiving at most {$N_\mathtt{err}$} erroneous messages for SFQ-based RM(1,3) encoder with $\pm$20\% PPV. Additionally, a no encoder design is simulated, which has only a 4-channel SFQ-to-DC converter interface (\textit{i.e.}, without any encoders and decoders). It is found that 1000 realizations are sufficient to produce relatively smooth and consistent CDF curves. As shown in Fig. \ref{fig:CDF_comparison_20PPV_v2}, the proposed RM(1,3) encoder has 86.7\% probability of transmitting {$N=100$} messages without any errors, whereas the no encoder design has 80.0\% probability. 
{One should note that the reported probability values in the CDF plots do not correspond to bit error rate but rather represent fabrication yield. Particularly, at $\pm$20\% PPV and $N_{err}=0$, the probability of 86.7\% corresponds to 867 out of 1000 circuit realizations (\textit{i.e.}, chips with specific PPV configuration) that have 100\% successful transmission rate without any bit errors.}

\begin{figure}[t]
	\centering
	\includegraphics[width=0.5\textwidth]{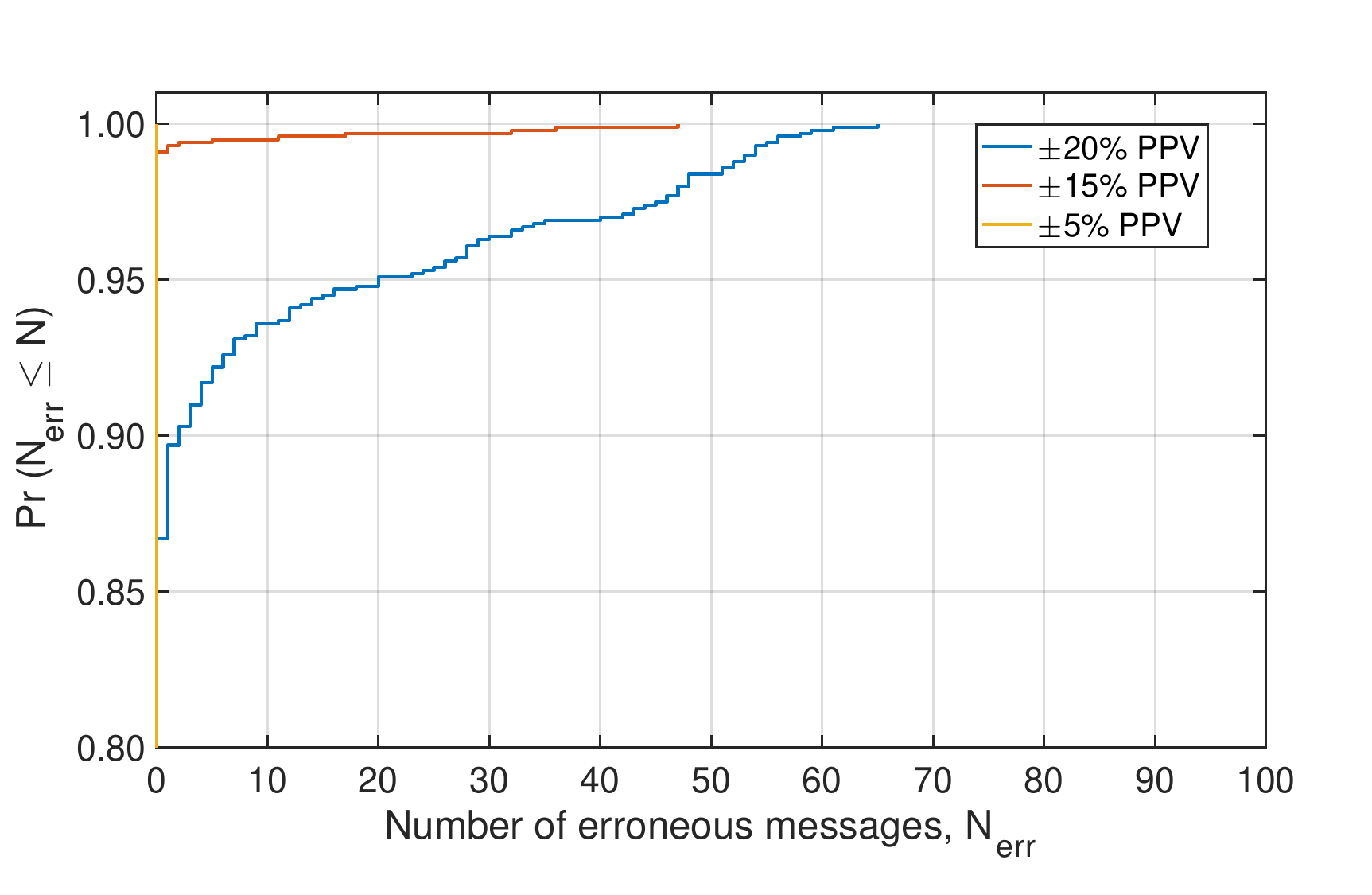}
	\caption{{CDF for receiving at most $N_\mathtt{err}$ erroneous messages with $\pm$20\%, $\pm$15\%, and $\pm$5\% PPV values for RM(1,3) encoder. $N=100$ messages are transmitted with 1000 independent circuit realizations.}}
	\label{fig:CDF_comparison_20_15_5_PPV}
\end{figure}

{The PPV of $\pm$20\%, which is presented in Fig. \ref{fig:CDF_comparison_20PPV_v2}, is considered to model extreme circuit parameter variations that can cause a significant number of errors. In more realistic cases, lower PPV values can be observed in modern superconductor fabrication processes \cite{tolpygo2014fabrication,tolpygo2014inductance,tolpygo2016advanced}. Fig. \ref{fig:CDF_comparison_20_15_5_PPV} shows the CDF plot for SFQ-based RM(1,3) encoder with $\pm$20\%, $\pm$15\%, and $\pm$5\% PPV values. Fig. \ref{fig:CDF_comparison_20_15_5_PPV} also shows that at $\pm$15\% and $\pm$5\% PPV, the RM(1,3) encoder corrects all errors with 99.1\% and 100\% probability, respectively.}

As discussed in Section \ref{section:RM_encoder_circuit}, the RM(1,3) encoder circuit consists of 49 cells (including XOR, DFF, splitter, and SFQ-to-DC converters). As compared to the no encoder design, which has only four cells (\textit{i.e.}, four SFQ-to-DC converters), the RM(1,3) encoder is more complex and prone to a higher chance of failure due to PPV. 
A similar pattern can be observed in Fig.~\ref{fig:CDF_comparison_20PPV_v2}. 
Before error-correction (see `RM(1,3): Before ECC' curve in Fig. \ref{fig:CDF_comparison_20PPV_v2}), there is roughly 55.4\% probability of having no errors in the received codewords. Therefore, it can be argued that even such lightweight encoder design introduces a significant amount of errors. Nevertheless, the error-correction capability of RM(1,3) encoder, where at most one bit-error can be corrected out of eight, offers better performance than the design without an encoder, as shown in Fig.~\ref{fig:CDF_comparison_20PPV_v2}.

\subsection{Effects of fabrication defects}\label{section:open_faults}

During the manufacturing process of SFQ chips, various defects may occur that can cause, \textit{e.g.}, open and/or short circuit faults. {Assuming that these types of faults are concentrated at the output terminals of interface circuits (\textit{e.g.}, damaged output pads/pins), the RM(1,3) encoder can operate correctly with at least seven out of eight channels.} Additionally, it can detect the errors with at least five out of eight channels. 

\begin{figure}[t]
	\centering
	\includegraphics[width=0.5\textwidth]{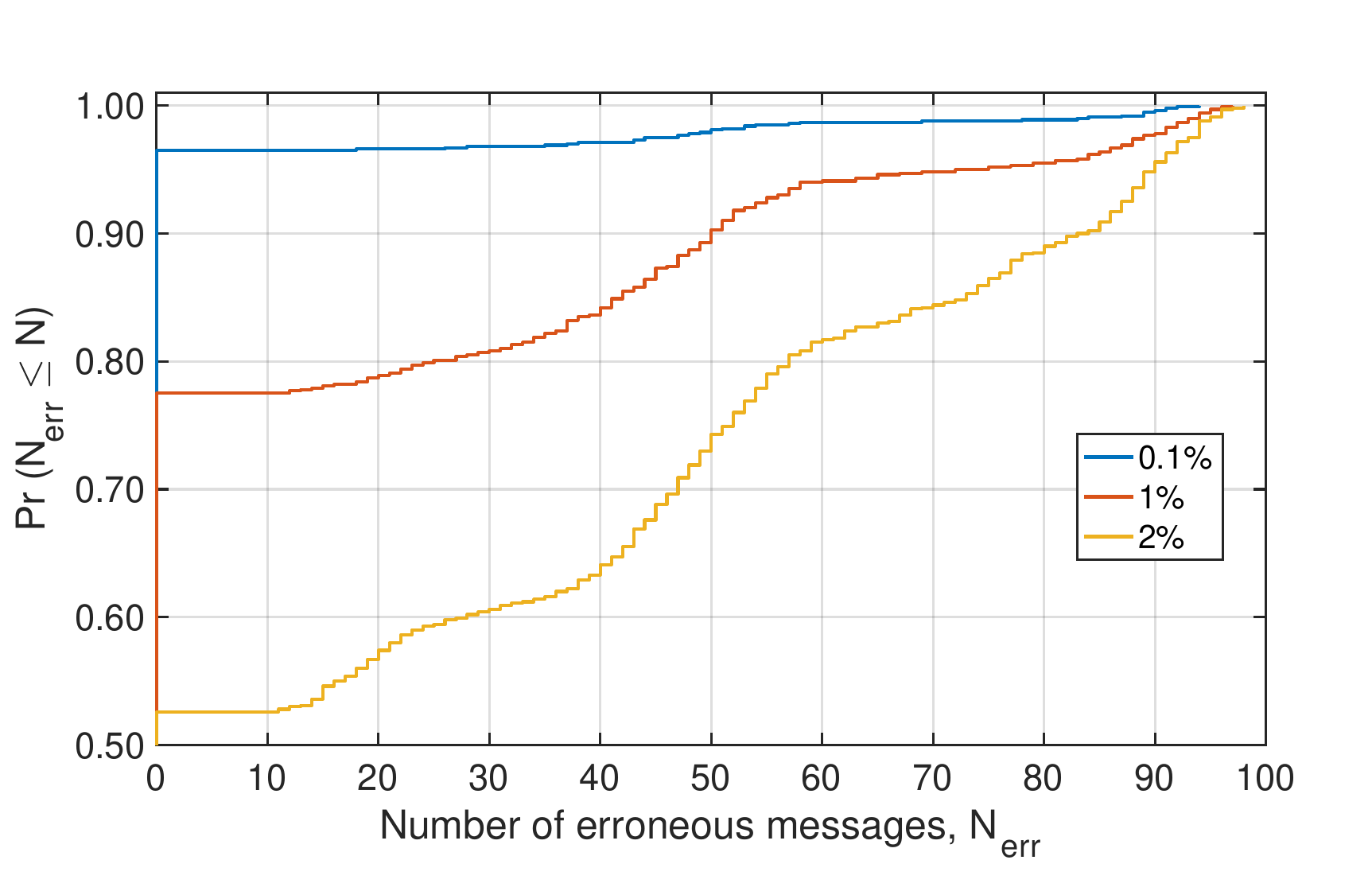}
	\caption{CDF for receiving at most {$N_\mathtt{err}$} erroneous RM(1,3) encoded messages. {$N=100$} messages are transmitted with 1000 independent circuit realizations. Each SFQ cell is subject to an open circuit fault with 0.1\%, 1\%, or 2\% probability.}
	\label{fig:CDF_comparison_open_circuit}
\end{figure}

{For more sophisticated analysis, assume that each SFQ cell has a certain fault probability.} 
{Since JoSIM lacks built-in fault analysis tools and functions, the proposed simulation framework (Fig. \ref{fig:framework_v2}) uses MATLAB to randomly select faulty cells and edit the JoSIM netlist files.} 
As a case study, let us consider the insertion of open circuit faults, where certain electrical path (\textit{e.g.}, wire) is broken and no signals can pass through it. In Fig. \ref{fig:framework_v2}, the MATLAB script chooses a predefined set of netlist line numbers, where each line corresponds to a subcircuit such as XOR gate, DFF, splitter, and SFQ-to-DC converter.
After that, it replaces the randomly selected lines with an empty string. 
{This creates an open circuit in which no SFQ pulses propagate to subsequent logic cells.}
Fig. \ref{fig:CDF_comparison_open_circuit} shows the CDF plot for receiving at most {$N_\mathtt{err}$} erroneous messages for RM(1,3) encoder with 0.1\%, 1\%, and 2\% open circuit fault probability. 
{Such analysis can serve as a figure of merit (FOM) for comparing multiple encoder designs.}

To maximize the tolerance of SFQ-based encoder to faults, it is important to account for its circuit structure and error-correction capability.
As mentioned previously, the RM(1,3) encoder can correct one bit error in the codeword.
{Reviewing the schematic of RM(1,3) encoder in Fig. \ref{fig:RM13_schematic_in_data_link} reveals that failures in two or more logic cells may still allow correct operation.} 
For example, for $c_8$, two DFFs could fail at the same time, leading to only one error in $c_8$ bit. Other similar pairs (namely, XOR and DFF cells) can be identified for $c_1$, $c_2$, and $c_7$. 
{The clock distribution network should be designed to connect these pairs through a common SFQ splitter cell.}
{Therefore, in the worst case scenario, failure of three cells (a pair of XOR and DFF, and the splitter connecting them) results in only one bit error, which the RM(1,3) encoder can correct. }






\section{Conclusion}\label{conclusion}

In this paper, a lightweight ECC encoder based on {the} RM(1,3) linear block code is proposed for SFQ-to-CMOS interface circuits. This encoder circuit is designed with SFQ logic cells. 
{A custom simulation framework using MATLAB and JoSIM is developed to evaluate the performance of RM(1,3) encoder.}
{Specifically, PPV and open circuit fault effects are studied using the proposed framework's automated data collection and analysis capabilities. 
The SFQ-based RM(1,3) encoder demonstrates higher error-free transmission probability than encoder-less design under $\pm$20\% or lower PPV.}


\bibliographystyle{./bibliography/IEEEtran}
\bibliography{./bibliography/IEEEabrv,./bibliography/IEEEexample}

\newpage
\onecolumn
\fontsize{20pt}{15pt}\selectfont

\end{document}